\documentclass[12pt]{article}
\usepackage{graphics,cite}

\def \gsim
{\raisebox{-3pt}{$\>\stackrel{>}{\scriptstyle\sim}\>$}}

\begin{document}
\begin{titlepage}
\begin{flushright}
CERN-PH-TH-2004-013\\
MPP-2004-10\\
SHEP-03-27\\
February 2004
\end{flushright}
\par \vspace{10mm}
\begin{center}
{\Large \bf
Matrix-Element Corrections \\[1ex]
to Parton Shower Simulations\\[2ex]
for Higgs Hadroproduction}
\end{center}
\par \vspace{2mm}
\begin{center}
{\large \bf G. Corcella$^{1,2}$ {\rm and} S. Moretti$^3$}\\
\vspace{2mm}
{$^1$Department of Physics, CERN\\ 
Theory Division\\
CH-1211 Geneva 23, Switzerland}\\
\vspace{2mm}
{$^2$
Max-Planck-Institut f\"ur Physik, Werner-Heisenberg-Institut,}\\
{F\"ohringer Ring 6, D-80805 M\"unchen, Germany}\\
\vspace{2mm}
{$^3$School of Physics and Astronomy, University of Southampton,\\
Highfield, Southampton, SO17 1BJ, UK}\\
\end{center}
\par \vspace{2mm}
\begin{center}
{\large \bf Abstract}
\end{center}
\begin{quote}
  \pretolerance 10000
We implement matrix-element corrections to HERWIG parton
shower simulations for Standard Model Higgs boson production at hadron 
colliders. We study the Higgs transverse momentum distribution and
compare different versions of HERWIG and resummed calculations.
The HERWIG results exhibit a remarkable improvement
as many more events are generated at large transverse momentum
after the inclusion of matrix-element corrections. 
\end{quote}
\end{titlepage}
\section{Introduction}
The Standard Model (SM) 
of electroweak interactions predicts the existence of the 
Higgs boson, which is responsible for the mechanism of mass generation.
However, such a particle has not been experimentally discovered yet.
Searches for the Higgs boson are one of the main goals of the
current experiments at the Tevatron accelerator and, ultimately, 
at future ones, like the Large Hadron Collider (LHC).

In order to perform such searches, precise QCD calculations are mandatory.
Other mechanisms may turn out to be useful at hadron colliders
\cite{review}, yet Higgs production via parton fusion is numerically
the most important one, especially at the LHC. Here, the leading-order 
processes in the strong coupling constant are
$q\bar q\to H$ and $gg\to H$, with the gluon-gluon fusion mechanism
overwhelming the quark-antiquark annihilation channel.

In order to investigate the phenomenology of the Higgs boson,
fixed-order calculations may be reliable as long as one only 
considers inclusive
observables, such as total cross sections; for less inclusive
quantities, one needs to account for multi-parton radiation
in order to perform trustworthy phenomenological analyses
\cite{joey}. 
Standard Monte Carlo (MC) algorithms \cite{herwig,pythia} describe parton
radiation in the soft or collinear approximation, but can have regions
of the phase space, so-called `dead zones', where no radiation is allowed.

In the dead zone, one can however 
rely on the fixed-order result, as in this region
the radiation is neither soft nor collinear enhanced. 
Several methods have recently been suggested in order to match parton showers 
and fixed-order matrix elements \cite{sey,sjo,frix,dobbs}. 
In this paper we follow the strategy which has been already used to
implement matrix-element corrections to the HERWIG event generator 
\cite{herwig} for several processes: 
$e^+e^-$ annihilation into quark pairs \cite{sey1}, 
Deep Inelastic Scattering (DIS) \cite{sey2},
top quark decay \cite{corsey1} and vector boson hadroproduction 
\cite{corsey2}.  The dead zone is here
populated by using the exact next-to-leading order (NLO)
tree-level matrix element result and the
shower in the already-populated region is corrected using the exact
amplitude any time an emission is capable of being the
hardest so far. 

In Section 2 we review the HERWIG parton shower algorithm for the
initial-state radiation. 
In Sections 3 and 4 we discuss the 
implementation of hard and soft matrix-element corrections, respectively.
In Section 5 we present results on the Higgs transverse momentum 
distributions using different versions of HERWIG and 
also   partonic  calculations from the literature. 
In Section 6 we summarize the main
results of our work. 

\section{The HERWIG parton shower algorithm}

HERWIG simulates the initial-state radiation in
hadron collisions according to a `backward evolution', 
in which the scale
is reduced away from the hard vertex
and traces the hard-scattering partons
back into the incoming hadrons \cite{marweb}. 
The algorithm relies on the universality of the elementary branching
probability in the soft or collinear approximation.
The probability of the emission
of an additional soft/collinear 
parton from a parton $i$ is given by:
\begin{equation}
  \label{elementary}
  dP={{dQ_i^2}\over{Q_i^2}}\;
  {{\alpha_S\left(\frac{1-z_i}{z_i}Q_i\right)}\over {2\pi}}\;
  P_{ab}(z_i)\; dz_i\;
  {{\Delta_{S,a}(Q^2_{i\mathrm{max}},Q_c^2)}\over{\Delta_{S,a}(Q_i^2,Q_c^2)}}\;
{{x_i/z_i}\over x_i}\;{{f_b(x_i/z_i,Q_i^2)}\over {f_a(x_i,Q_i^2)}}.
\end{equation}
The HERWIG ordering variable is
$Q_i^2=E^2\xi_i$, where $E$ is the energy of the parton that splits
and $\xi_i={{p\cdot p_i}\over{E E_i}}$, with $p$ and $p_i$ being the
four-momenta of the splitting and of the emitted parton, respectively;
$z_i$ is the energy fraction of the outgoing space-like
parton with respect to the incoming one; $P_{ab}(z)$ is the Altarelli--Parisi
splitting function for a parton $a$ evolving in $b$.
In the massless approximation, 
$\xi_i=1-\cos\theta$, where $\theta$ is the emission angle to the
incoming-hadron direction. 
For soft emission ($E_i\ll E$),  ordering according to $Q_i^2$
corresponds to angular ordering. 
In (\ref{elementary}) $f_a(x_i,Q_i^2)$ is the parton distribution function
for the partons of type $a$ in the initial-state hadron, $x_i$ being the
parton energy fraction.
The quantity
\begin{equation}
\Delta_S(Q_2^2,Q_1^2)=\exp\left[-{{\alpha_S}\over {2\pi}}
\int_{Q_1^2}^{Q_2^2}{{dk^2}\over{k^2}}\int_{Q_1/Q_2}^{1-Q_1/Q_2}
{dzP(z)}\right] \end{equation}
is the Sudakov form factor, expressing the probability of evolution 
from $Q_2^2$ to $Q_1^2$ with no resolvable branching. 
Unitarity dictates that the Sudakov form factor
sums up all-order virtual and unresolved real contributions. 
In (\ref{elementary}), $Q_{i\mathrm{max}}$ is the maximum value of $Q$, fixed
by the hard process, and $Q_c$ is the value 
at which the backward evolution is terminated, corresponding, in the case of 
HERWIG, to a cutoff on the transverse momentum
of the showering partons. 
However, since $Q_c$ is smaller than the minimum scale at which 
the parton distribution functions are evaluated, an additional cutoff 
on the evolution variable $Q_i$ has to be set.
If the backward evolution has not resulted in a
valence quark, an additional non-perturbative parton emission is generated to
evolve back to a valence quark. Such a valence quark has a Gaussian
distribution with respect to the non-perturbative intrinsic transverse momentum
in the hadron, with a width $q_{T{\mathrm{int}}}$,  
that is an adjustable parameter and whose default value is zero.

As the variables $Q_i^2$ and $z_i$ are frame-dependent, one needs to
specify the frame where the evolution occurs.
One can show that, as a result of the 
$Q^2$-ordering, the maximum $Q$-values of two colour-connected partons
$i$ and $j$ having momenta $p_i$ and $p_j$ 
are related via $Q_{i\mathrm{max}}Q_{j\mathrm{max}}=p_i\cdot p_j$,
which is Lorentz-invariant.
For Higgs boson production in hadron collisions, 
symmetric limits are set in HERWIG:
$Q_{i\mathrm{max}}=Q_{j\mathrm{max}}=\sqrt{p_i\cdot p_j}$.
Furthermore, the energy of
the parton which initiates the cascade is set to
$E=Q_{\mathrm{max}}=\sqrt{p_i\cdot p_j}$.
Such conditions define the HERWIG frame.
It follows that ordering according to $Q^2$ implies that, in the
showering frame, $\xi<z^2$.

The region $\xi>z^2$ is therefore a `dead zone' for the shower 
evolution.
In such a zone the physical radiation is not soft or collinear
enhanced, but not completely absent as it happens in the 
standard algorithm.
It is indeed the purpose of this paper to improve the HERWIG parton
shower algorithm and populate the dead zone 
by including matrix-element corrections.

\section{Hard matrix-element corrections}
The Born processes leading to Higgs production via parton fusion
at hadron colliders are
$gg\to H$ (gluon-gluon fusion), of ${\cal O} (\alpha_S^2\alpha)$, 
and $q\bar q\to H$ (quark-antiquark annihilation), of ${\cal O}(\alpha)$.
In the SM, Higgs production via
gluon-gluon fusion is mediated by a quark loop. 
Hereafter, we shall 
consider only the top quark contribution in the loop, which is 
largely dominant in the SM, with finite top mass.

Gluon-gluon fusion receives next-to-leading order (NLO)
corrections of ${\cal O}(\alpha_S^3\alpha)$ due to the
elementary processes $gg\to gH$, $gq\to qH$, $qg\to qH$ and $q\bar q\to gH$.
The corrections to $q\bar q\to H$ are instead $q\bar q\to gH$, 
$qg\to qH$ and $gq\to qH$\footnote{We point out that processes
$q\bar q\to gH$ and $qg\to Hq$ are NLO corrections to both 
gluon-gluon fusion and quark-antiquark annihilation. 
However, the actual matrix elements
are different whether they are corrections to $gg\to H$, i.e., at
one-loop, or to $q\bar q\to H$, i.e., at tree-level.},
of ${\cal O}(\alpha_S\alpha)$.

The matrix elements squared for the corrections to $gg\to H$ can be found in
\cite{bauglo,spira}, where top mass effects are fully included, and in
\cite{ellis}, where the authors have considered the 
infinite top mass approximation in the loop. HERWIG uses the formulae
of Ref.~\cite{bauglo}.
The actual expressions for such amplitudes are rather involved once
the top mass is fully taken into account; therefore,
we do not report them here for the sake of brevity. 
The real NLO corrections to $q\bar q\to H$ are instead rather straightforward:
the formulae we use can be read from Eq.~(3.62) of \cite{MSSM},
with appropriate Yukawa couplings and crossing.

In order to implement matrix-element corrections to Higgs production,
we follow the prescription of Ref.~\cite{sey3}, where a method to
include higher-order corrections from real radiation
to a generator of the lowest-order process has been proposed.

Referring, e.g., to the correction to gluon-gluon fusion given by
\begin{equation}
g(p_1)g(p_2)\to g(p_3)H(q),
\label{gggh}
\end{equation}
we define the partonic Mandelstam variables of process (\ref{gggh})
\begin{equation}
\hat s=(p_1+p_2)^2,\ \hat t=(p_1-p_3)^2,\ \hat u=(p_2-p_3)^2.
\end{equation}
As in \cite{sey3}, we consider a generic Higgs decay $H\to \ell_1\ell_2$ and 
compute the differential cross sections $d\sigma_3$ for
the $2\to 3$ process $gg\to g\ell_1\ell_2$ and $d\sigma_2$ for the $2\to 2$ one
$gg\to \ell_1\ell_2$ via an intermediate Higgs boson.

As done in \cite{corsey2}, if we assume that the Higgs squared momentum and
rapidity in the Born process are conserved in the transition
from the $2\to 2$ to the
$2\to 3$ process, we find that a factorization formula holds.
We obtain:
\begin{equation}
{{d\sigma_3}\over{d\sigma_2}}=
{1\over{16\pi\sigma_0m_H^2}} {{d\hat sd\hat t}\over{\hat s^2}} 
|{\cal M}(gg\to gH)|^2 {{f_1(\xi_1,m_T^2)f_2(\xi_2,m_T^2)}
\over{f_1(\chi_1,m_H^2)f_2(\chi_2,m_H^2)}},
\label{ds3}
\end{equation}
where ${\cal M}$ is the matrix element for the
process of Eq.~(\ref{gggh}).
In Eq.~(\ref{ds3}) $\sigma_0$ is the cross section of the
Born process $gg\to H$,
$f_1$ and $f_2$ are the parton distribution functions
of the hard-scattering partons in the incoming hadrons, $\xi_i$ and 
$\chi_i$ are the energy fractions of the incoming partons in 
the $2\to 3$ and $2\to 2$ processes, respectively.

For the $gg\to H$ processes we evaluate the parton distribution functions
and the strong coupling constant at $m_H^2$, which is the hard scale 
of the $gg, q\bar q\to H$ processes.
For the $gg\to Hg$ case, we set such scales to the Higgs transverse mass, 
$m_T^2=q_T^2+m_H^2$, $q_T$ being the $H$ transverse momentum,
which accounts for the additional parton emission 
via matrix-element corrections.
Such choices are the same as the ones done for $W/Z$ production
\cite{corsey2}.

To include matrix-element corrections, one will have to implement
the weight function given by Eq.~(\ref{gggh}).
In order to get the phase space which is populated by HERWIG and 
the dead zone, where no radiation is allowed, we can repeat all the
steps which have been employed for Drell--Yan interactions
and report here just 
the final results obtained in \cite{corsey2}.
The total phase-space limits, in terms of the variables $\hat s$ and 
$\hat t$, read:
\begin{eqnarray}
\label{tots} m_H^2 <&\hat s&<s, \\ 
\label{tott} m_H^2-\hat s <&\hat t&<0,
\end{eqnarray}
where $s$ is the hadronic centre-of-mass energy squared.
The value $\hat s=m_H^2$ corresponds to soft radiation,
and the lines $\hat t=0$ and $\hat t=m_H^2-\hat s$ to collinear
emission.

As shown in Ref.~\cite{corsey2}, the HERWIG phase space, which corresponds
to $\xi<z^2$ in terms of the showering variables, is given by:
\begin{eqnarray}
{{m_H^2}\over 2} (7-\sqrt{17}) <&\hat s &<s\\
\hat t_{\mathrm{min}}< &\hat t &< -{{m_H^2}\over 2}
\left(3-\sqrt{1+{{8m_H^2}\over{\hat s}}}\right),
\label{herph}
\end{eqnarray}
with $\hat t_{\mathrm{min}}=m_H^2-\hat s-\hat t_{\mathrm{max}}$.
\begin{figure}
\centerline{\resizebox{0.65\textwidth}{!}{\includegraphics{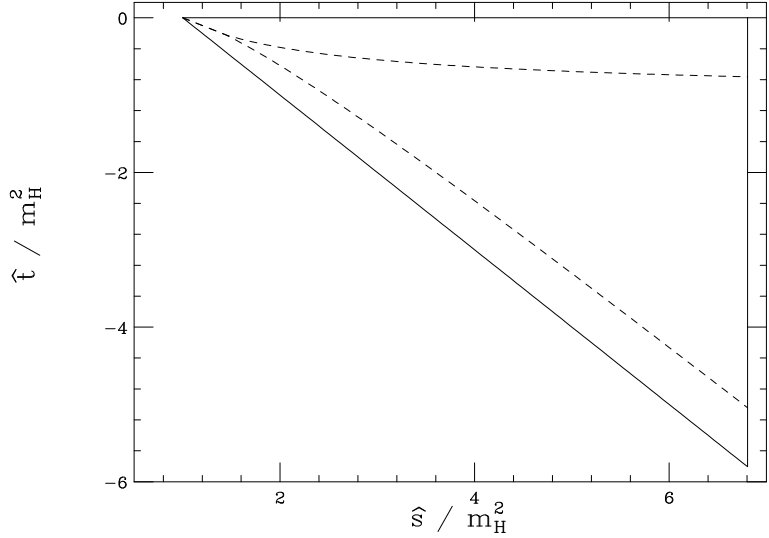}}}
\caption{\small Total (solid) and HERWIG (dashed) phase space for 
a Higgs mass $m_H=115$~GeV and centre-of-mass energy 
$\sqrt{s}=300$~GeV.}
\label{phase}
\end{figure}
In Fig.~\ref{phase} it is plotted the physical phase space, along 
with the region which the showering algorithm populates, for 
$m_H=115$~GeV, which is the HERWIG default value,
and $\sqrt{s}=300$~GeV. The soft and collinear regions
are covered by the shower and one has an overlapping region, where radiation
may come from either parton. 
As expected, Fig.~\ref{phase} exhibits the presence of a dead zone, 
where the standard HERWIG algorithm allows no radiation.

Similar arguments are also valid for the other corrections to the $gg\to H$ 
Born process. Nevertheless, a few comments are in order.
The processes $qg\to qH$ and $gq\to qH$ are not soft divergent, but only
collinear. This implies that the lower limit for, e.g., $gq\to qH$ is
$\hat t_{\mathrm{min}}=m_H^2-\hat s$ and there is no overlapping region.
Likewise, for $qg\to qH$, the upper limit is $\hat t_{\mathrm{max}}=0$. 
The process $q\bar q\to gH$, when taken as a correction to $gg\to H$,
 cannot be interpreted
in the parton-shower language as it is neither soft nor collinear
singular. However, we included this process as well via matrix-element
corrections: the dead zone will then be the full physical phase space
given by Eqs.~(\ref{tots})--(\ref{tott}).

In order to implement hard matrix-element corrections, we populate the
dead 
zone using the probability distribution given by the exact matrix element,
i.e., Eq.~(\ref{ds3}), where ${\cal M}$ will have to be the 
correct amplitude squared of the hard-scattering process.
Moreover, in order to maximize the efficiency of the event generation,
the fraction of events generated according to the different
subprocesses $gg$, $qg$, $gq$ or $q\bar q$ is proportional to the
corresponding cross sections in the so-called `$H$ + jets' process, where the
hard process is always one of the corrections to $gg\to H$.

Hard corrections to $q\bar q\to H$ processes are similar to what discussed
for the gluon-gluon fusion channel. 
In particular, the phase-space configuration for $q\bar q\to gH$ is
as in Fig.~\ref{phase}; the one for $qg\to qH$ or $gq\to gH$ is the 
same as for the analogous corrections to gluon-gluon fusion. 

\section{Soft matrix-element corrections}
The implementation of soft-matrix-element corrections can be performed 
using the general method of Ref.~\cite{sey}. Any time an emission in the
HERWIG phase space is capable of being the hardest so far, we use the
exact matrix element instead of the standard HERWIG algorithm.
The hardness of a radiation is measured in terms of the transverse
momentum of the emitted parton with respect to the emitting one.

The inclusion of the soft correction is performed by multiplying the
parton shower branching probability by a factor which is the ratio of
the HERWIG to the matrix-element distribution.
It reads:
\begin{equation}
{{d^2\sigma}\over{d\hat sd\hat t}}={{d^2\sigma}\over{d zd\xi}}
J(\hat s,\hat t;z,\xi).
\label{softme}
\end{equation}
In Eq.~(\ref{softme}), $J(\hat s,\hat t;z,\xi)$ is the Jacobian factor for the
transformation $(z,\xi)\to (\hat s,\hat t)$. 
As the kinematics for Higgs production are the same as for vector boson
production, we can just use for the Jacobian factor the result reported
in Ref.~\cite{corsey2}.

Before closing this section, we would like to point out that matrix-element
corrections to Higgs production are differently implemented
in the PYTHIA event generator \cite{pythia}.
In fact, in PYTHIA the parton shower approximation is used in all
the physical phase space and the exact matrix element corrects only the
first emission, rather than the hardest-so-far one.
Furthermore, the approximation of a top quark of infinite
mass is used in PYTHIA to define the ratios of the $gg\to Hg$ and $qg\to Hq$ to
$gg\to H$ cross sections, while the latter does use the complete 
expressions\footnote{Also, corrections to the $q\bar q$ subprocesses are not 
available in PYTHIA.}.
A comparison of phenomenological results obtained running HERWIG
and PYTHIA is given in \cite{joey,LH2003}.

\section{Transverse momentum distribution}
We would now like to present results for the Higgs transverse momentum
distribution and investigate the impact of matrix-element corrections.
In particular, we wish to compare HERWIG results without and with such

corrections as well as the improved HERWIG version with the
resummed calculation of Ref.~\cite{bozzi} and the so-called `Monte Carlo at 
next-to-leading order' (MC@NLO) implementation
of \cite{mc}.
We shall consider Higgs production at the Tevatron and LHC and
we shall always assume that the intrinsic transverse momentum is
$q_{T,\mathrm{int}}=0$. Unless otherwise stated, 
we shall use the default parton distribution 
functions of HERWIG,
but we can anticipate that the relative effect
of matrix-element corrections is basically the same, regardless of the
chosen set of parton distribution functions.
\begin{figure}
\centerline{\resizebox{0.65\textwidth}{!}{\includegraphics{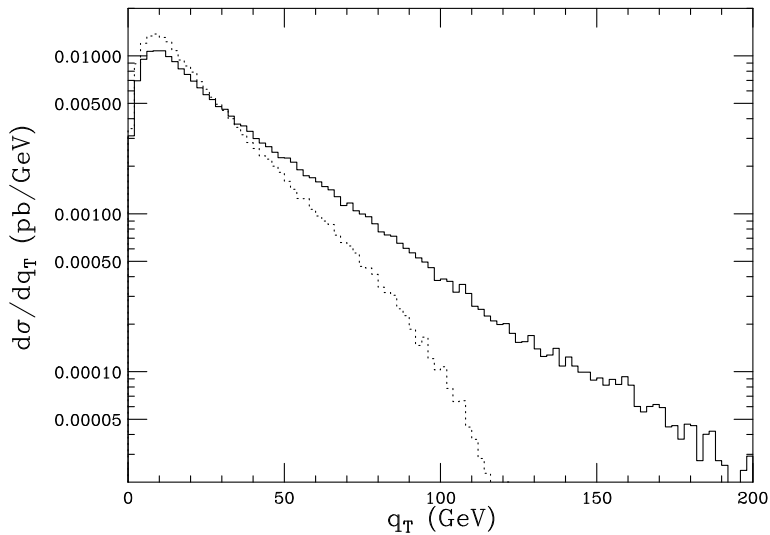}}}
\caption{\small Higgs transverse momentum distribution at the Tevatron, 
according to HERWIG with (solid) and without (dotted) 
matrix-element corrections.
We have set the Higgs mass to $m_H=115$~GeV.}
\label{tev}
\end{figure}
\begin{figure}
\centerline{\resizebox{0.65\textwidth}{!}{\includegraphics{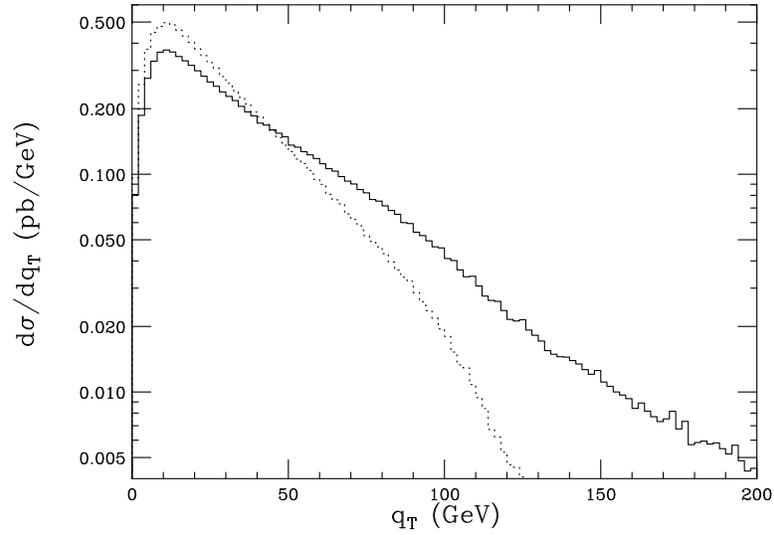}}}
\caption{\small As in Fig.~\ref{tev}, but at the LHC.}
\label{lhc}
\end{figure}

In Fig.~\ref{tev} we consider Higgs production at the Tevatron, 
i.e., $p\bar p$ collisions at $\sqrt{s}=2$~TeV, which is the 
centre-of-mass energy of the current Run 2.
In Fig.~\ref{lhc} we plot instead the $q_T$ distribution at the LHC, i.e.
$pp$ collisions at $\sqrt{s}=14$~GeV. 
We consider the HERWIG prediction with (solid histogram) and without (dotted
histogram)
matrix-element corrections. Beyond $q_T\simeq m_H/2$ the matrix-element 
corrected version allows for many more events. In fact, one can prove 
that, within the standard algorithm, $q_T$ is constrained to be
$q_T<m_H$: the events at large $q_T$ are therefore generated via the same
exact amplitude. At small $q_T$, the prediction which includes hard and soft
corrections displays a suppression. By default, after matrix-element 
corrections, the total normalization is still the same and equal
to the LO result. It is therefore reasonable that the enhancement
at large $q_T$ implies a reduction of the number of
events which are generated at small transverse momentum.
This result was already found for $W/Z$ production \cite{corsey2} and
is visible especially at the LHC.

\begin{figure}
\centerline{\resizebox{0.65\textwidth}{!}{\includegraphics{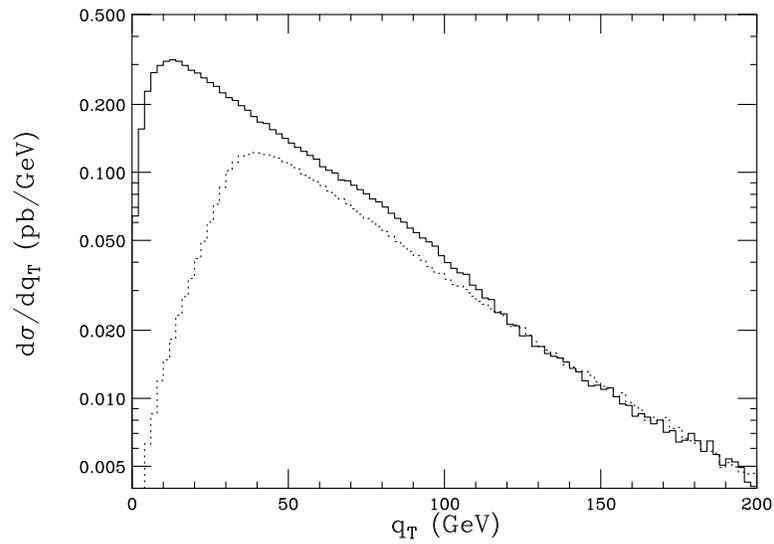}}}
\caption{\small Comparison of matrix-element corrected HERWIG
prediction (solid) and `$H$ + jets' (dotted) at the LHC. Here,
$q\bar q\to H$ processes have been turned off.}
\label{lhc1}
\end{figure}
\begin{figure}
\centerline{\resizebox{0.65\textwidth}{!}{\includegraphics{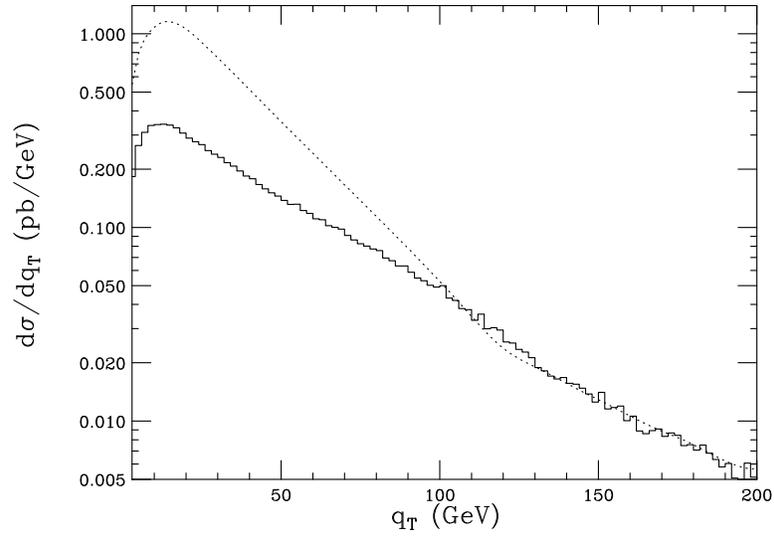}}}
\caption{\small Comparison of matrix-element corrected HERWIG
prediction (solid) and the resummed calculation of [\ref{bozzi}]
(dotted).}
\label{resum}
\end{figure}
\begin{figure}
\centerline{\resizebox{0.65\textwidth}{!}{\includegraphics{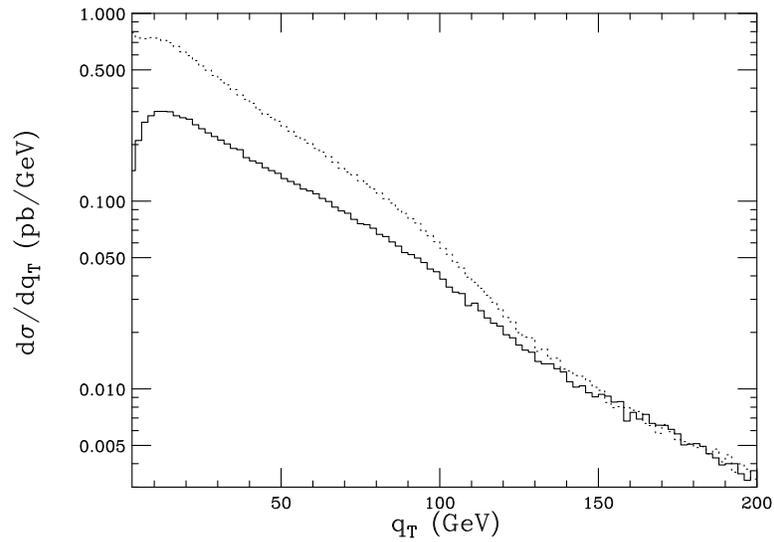}}}
\caption{\small Comparison of matrix-element corrected HERWIG
(solid) and MC@NLO (dotted) predictions.}
\label{mc1}
\end{figure}
In Fig.~\ref{lhc1} we plot the improved HERWIG spectrum (solid line) for the 
LHC,
along with the result obtained running the 
$H$ + jets process (dotted line).
In order for such a comparison to be reliable, we have turned 
the $q\bar q\to H$ hard process off, as `$H$ + jets' does not currently
implement the corrections to quark-antiquark annihilation.
We use the cutoff $q_{T,\mathrm{min}}=30$~GeV for the `$H$ + jets' 
generation.
Fig.~\ref{lhc1} shows that at small $q_T$ the two predictions are
fairly different, but at large transverse momentum they 
agree.
This is a reasonable result, since, after matrix-element corrections,
large-$q_T$ events of both processes are generated via the exact amplitude.
This is a check that the implementation of the hard correction is
reliable.

Next we compare the new HERWIG version with the
resummed partonic calculation of Ref.~\cite{bozzi}, where the
authors resummed soft higher-order contributions to the $gg\to H$ 
process.
In fact, the 
differential distribution $d\sigma/dq_T$ 
for the production of a Higgs boson of transverse momentum $q_T$
presents terms $\sim 1/q_T^2\  \alpha_S^n\ln^m(m_H^2/q_T^2)$
which get arbitrarily large at small $q_T$, i.e., for
$q_T\ll m_H$.
The leading
logarithms (LL) correspond to $m=n+1$, the next-to-leading 
logarithms (NLL) to $m=n$, the next-to-next-to-leading logarithms 
(NNLL) to $m=n-1$.

The authors of Ref.~\cite{bozzi} have resummed such enhanced logarithms to
NNLL level at small $q_T$
and matched them to the NNLO result at large $q_T$
in order to
obtain a reliable result over the full $q_T$ range. However, for the
sake of comparison with HERWIG, which includes leading logarithms and
only some subleading terms (see, e.g., the discussion in Ref.~\cite{cor}
on the comparison of HERWIG and resummation for $W/Z$ production), 
we use the results
of \cite{bozzi} in the NLL approximation, matched to the NLO 
prediction\footnote{We point out that the NLO corrections to 
$gg\to H$ for the
cross section, of order ${\cal O} (\alpha_S^3\alpha)$, are actually LO for 
the $q_T$ distribution and according to the notation of Ref.~\cite{bozzi}.
For the sake of self-consistency, we stick to our conventions and call
NLO the fixed-order calculation to which the resummation is matched.}.

In order for such a comparison to be
trustworthy, we have to do the same assumptions as \cite{bozzi}.
We use a Higgs mass value $m_H=125$~GeV
and, once we apply matrix-element corrections, in Eq.~(\ref{ds3}) 
we evaluate the strong coupling constant and the parton distribution 
functions at a scale given by the Higgs mass $m_H^2$.
As in \cite{bozzi}, we also use the approximation of a top quark with infinite
mass in the
loop, which is a user-defined option of the HERWIG event generator,
and choose the MRST2001 leading-order parton
distribution functions \cite{mrst2001}.
We finally turn the Born $q\bar q$-initiated processes off.


The result of the comparison is shown in Fig.~\ref{resum}.
The normalization and the small-$q_T$ behaviour are clearly different. In
fact, the total cross section is LO in HERWIG and NLO in our
approximation of the work in
\cite{bozzi}. The discrepancy at small transverse momenta is instead due
to the different logarithmic accuracy which the two considered approaches 
implement.
However, the two curves agree at large $q_T$, where the NLO calculation
dominates. This is another independent check of the reliability of
the implementation of matrix-element corrections.

Finally, 
we would like to compare the results of standard HERWIG after matrix-element
corrections with the MC@NLO event generator 
(version 2.2) \cite{mc}, 
which implements numerically the method of 
Ref.~\cite{frix} to simulate Higgs boson production at hadron colliders.
As discussed in \cite{frix}, the MC@NLO approach implements both real
and virtual corrections to the hard-scattering process, in such
a way that predicted observables, like total cross sections,
are correct to NLO accuracy. Moreover, the MC@NLO showers turn
soft matrix-element corrections off.

Version 2.2 of the MC@NLO~\cite{mc} currently includes only the corrections
to Higgs production in the gluon-gluon fusion channel, hence we shall again
have to turn the quark-annihilation processes off for the sake of a
reliable comparison.
In both matrix-element corrected HERWIG and MC@NLO generators
we set factorization and renormalization scales for the 
NLO processes like $gg\to Hg$ equal to the Higgs transverse mass.

In Fig.~\ref{mc1} we show the result of this comparison, which
exhibits similar features to the one with the resummed calculation.
The normalization and
the small-$q_T$ behaviour of the two curves are different, but
nevertheless the large transverse momentum predictions are in 
good agreement.

\section{Conclusions}
We have implemented matrix-element corrections to HERWIG
parton shower simulations for direct Higgs production at hadron colliders. 
We have considered corrections to 
$gg\to H$ and $q\bar q\to H$ and have used the exact 
tree-level NLO matrix element to
populate the HERWIG dead zone of the physical phase space and for every 
hardest-so-far emission in the already-populated region.

We have considered the Higgs transverse momentum $q_T$ distribution and
have compared HERWIG predictions with and without matrix-element
corrections. We have found a remarkable effect of such corrections
at large $q_T$, as many more events are generated here
via the exact amplitude. 
We have compared the matrix-element corrected result with the
$q_T$ prediction yielded by the HERWIG `$H$ + jets' process and have found
agreement at large transverse momentum. 

We have compared HERWIG provided with hard and soft corrections
with a recent NLL+NLO soft-resummation calculation and, after making 
consistent
choices for the Higgs masses and the scales entering the calculation, we have
found very good agreement in the large-$q_T$ range.

Moreover, the HERWIG implementation with NLO real corrections has
feared rather well against the MC@NLO result, as obtained by
using both real and virtual NLO corrections to the hard partonic process.
Besides obvious differences in the total normalization
(which is LO in standard HERWIG and NLO in the MC@NLO approach) and
at small $q_T$, the large-$q_T$ spectra agree well, which
is another consistency check of the reliable inclusion of matrix-element
corrections.

Between the described implementation and the one available within the 
MC@NLO option,
we believe that HERWIG is presently a reliable event generator for
Higgs production at hadron colliders both at small and large transverse
momentum and that the currently-available
matrix-element corrections will play an important role to perform any
analysis on Higgs searches at present and future colliders. In particular,
the option described here may be the most convenient choice for
transverse momentum values $q_T\gsim m_H$.

Finally, it should be noted that the HERWIG implementation presented in this 
paper includes both a finite value for the quark mass in the loop of the
relevant gluon-gluon fusion subprocesses 
and the corrections to the quark-antiquark
annihilation channels,
thus lending itself to a generalisation of the algorithm
to the case of the Minimal Supersymmetric Standard Model, work
which is currently in progress.

\section*{Acknowledgements}
We are grateful to S. Frixione for his help in the
use of the MC@NLO program and for many useful suggestions and comments.
We are indebted to M.H. Seymour 
for several discussions on the topic of matrix-element
corrections.
We also thank S. Catani, M. Grazzini and B.R. Webber
for useful conversations and plots.

\end{document}